\documentstyle[aps,twocolumn,epsfig,amssymb,amstex]{revtex}

\newcommand{\curl}{\operatorname{curl}}
\newcommand{\dv}{\operatorname{div}}

\renewcommand{\exp}{\operatorname{exp}}

\newcommand{\lapl}{\operatorname{\Delta}}

\newcommand{\undertilde}[1]{\mbox{$\underset{\displaystyle@!@!@!@!\widetilde{}}{#1}$}}
\newcommand{\ovec}[1]{{\mbox{\boldmath $#1$}}}
\newcommand{\xvec}{\ovec{x}}
\newcommand{\kvec}{\ovec{k}}

\newcommand{\evec}{\ovec{e}}

\newcommand{\ezvec}{\ovec{e}_z}
\newcommand{\zervec}{\ovec{0}}
\newcommand{\tmax}{{\text{max}}}
\newcommand{\Bvec}{\ovec{B}}
\newcommand{\bvec}{\ovec{b}}
\newcommand{\eBvec}{\ovec{e}_B}

\newcommand{\jvec}{\ovec{j}}
\newcommand{\omvec}{{\ovec{\omega}}}

\renewcommand{\ovec}{\vec}

\begin{document}

\title{A Hall--Drift Induced  Magnetic Field Instability}

\author{M. Rheinhardt and U. Geppert}
\address{Astrophysikalisches Institut Potsdam,
         An der Sternwarte 16, 14482, Potsdam, GERMANY}

\date{\today}

\maketitle

\begin{abstract}
In the presence of a strong magnetic field and under conditions as
realized in the crust and the superfluid core of neutron stars the Hall--drift
dominates the field evolution.
We show by a linear analysis that for a sufficiently strong large--scale
background field depending at least quadratically on position
in a plane conducting slab an instability occurs which rapidly generates
small--scale fields. Their growth rates depend on the choice of the boundary
conditions, increase with the
background field strength and may reach $10^3$ times the ohmic decay rate.
The effect of that instability on
the rotational and thermal evolution  of neutron stars is discussed.
\end{abstract}


In the presence of a magnetic field the electric conductivity
becomes a tensor and, what is more, 
two non--linear effects are introduced into Ohm's law:
the Hall--drift and the ambipolar diffusion.
However, if the conducting matter consists of electrons and one sort of 
ions and no neutral particles take part in the transport processes
the ambipolar diffusion is absent \cite{YS91}. Such a
situation is realized e.g. in crystallized crusts of neutron stars
and/or in their cores if the neutrons are superfluid,
but the protons are normal and the
electrons may therefore collide with protons but effectively not with the
neutrons.

The effect of the Hall--drift on the magnetic field evolution
of isolated neutron stars has been considered by a number of authors (see
e.g. \cite{HUY90SU97,GR92,M94,NK94US99,SU95,US95,VCO99}). 
They discussed the
redistribution of magnetic energy from an initially
large--scale (e.g. dipolar) field into small--scale
components due to the non--linear 
Hall--term. Though the Hall--drift is a non--dissipative process,
the tendency to redistribute the magnetic 
energy into small scales may accelerate the field decay considerably.

Indeed, when starting with a large--scale magnetic field
the {\it Hall--cascade} derived in \cite{GR92} will generate small--scale field
components down to a scalelength $l_{crit}$, where the ohmic dissipation begins
to dominate
the Hall--drift. Considering numerically the evolution of a magnetic field
in a sphere consisting of spherical harmonics up to a multipolarity $l=5$,
Shalybkov \& Urpin \cite{SU95}
concluded that the inclusion of higher harmonics will not influence the
magnetic
evolution. This conclusion is what we want to put in question.

In some of the above--mentioned investigations
numerical instabilities are reported if either too many harmonics were taken
into account \cite{US95} or the initial field is too strong \cite{NK94US99}.
Also, when considering the thermomagnetic field generation in the crust
of young neutron stars \cite{WG96}, where
small--scale modes are the first ones to be excited 
numerical instabilities occurred caused exclusively by the Hall--drift.

Here we want to show that all the observed instabilities are very likely
in their essence not of
numerical origin but have physical reasons. In presuming that we felt strongly
supported by the
close analogy of the linearized field evolution equation including Hall--drift
to the induction
equation including the so--called $\omvec\times \jvec$--effect introduced by 
R\"adler \cite{R69}. Within the framework of mean--field dynamo theory
(see e.g. \cite{M78}) he demonstrated the possible occurrence of magnetic
instabilities
in an electrically conducting fluid if a shear flow acts together with an
electromotive force (e.m.f.)
perpendicular to the current density $\jvec \propto \curl\Bvec$.

To prove the existence of a {\it Hall--drift induced instability} we 
employ a simplified model: We assume spatial constancy
of the conductive properties of the matter
and show that under special
conditions with respect to the (large--scale) background field strength and
geometry 
an instability occurs which quickly transfers magnetic energy from the
background field to small--scale perturbations.
We present 
the result of a linearized analysis which returns only growth 
rates and the spatial structure of the 
unstable field modes. Only a fully non--linear analysis is able to yield
saturation values of the excited small--scale modes.

The instability may work in different physical systems 
but is probably most efficient in modifying the field decay in compact
astrophysical bodies. 
Then it will act only during an episode of the field decay, which
unavoidably
leads to a zero field. This episode, however, may have observable
consequences. 


In the absence of motions and of ambipolar diffusion,
the equations which govern the magnetic field are
\begin{equation}
\hspace*{-6mm}\begin{aligned}
&\dot{\Bvec} = -  c~~\curl\bigg(\frac{c}{4\pi\sigma}
\big(\curl\Bvec+ \omega_B\tau_e\,(\,\curl\Bvec
\times\vec e_B\,)\big)\!\bigg)\\
&\dv \Bvec = 0\;,
\end{aligned}
\hspace*{-5mm}\label{indeq}
\end{equation}
where $c$ is the speed of light, $\sigma$ the electric conductivity
caused by electrons, $\tau_e$ the electron relaxation
time and $\omega_B= e|\Bvec|/m_e^* c$ the electron
Larmor frequency, with $e$ being the elementary charge and $m_e^*$ the
effective mass of an electron. $\eBvec$ is the unit vector
in $\Bvec$--direction. An estimate of the two terms on the r.h.s. of
\eqref{indeq} gives rise to the supposition that the
Hall--drift becomes important only if $\omega_B\tau_e > 1$.

Using standard arguments one can immediately state that in the absence
of currents at infinity the total energy of any
solution of \eqref{indeq} is bound to decrease monotonically to zero since the
Hall--term $\propto\curl\Bvec\times\evec_B$ is unable to deliver energy
(nor to consume it).

For simplicity we assume the conductive properties of the matter to be
constant in space and time, that is, we assume constant
$\sigma$ and $\tau_e/m_e^*$. Thus, the 
induction equation can be rewritten in dimensionless variables such that
it no longer contains any parameter and the magnetic field evolution is solely
determined by its initial configuration $\Bvec(\xvec,0)$.
For that purpose we normalize the
spatial coordinates by a characteristic length $L$ of the model
(for a neutron star it could be, e.g., its radius or the thickness of its
crust), the time by the ohmic decay time $4\pi\sigma L^2/c^2$ and the
magnetic field by $B_N=m_e^* c/ e\tau_e$. The
governing equations in these dimensionless variables read 
\begin{equation}
\dot{\Bvec} = \Delta\Bvec - \curl (\,\curl\Bvec
\times\Bvec\,)\,\;,\;\;\dv \Bvec = 0\;, 
\label{indeqdimless}
\end{equation}
where the differential operations have to be performed with respect to the
now dimensionless spatial and time coordinates
$x,y,z$ and $\tau$, respectively.

Stepping now into the search for instabilities we first have to define a
proper reference state $\Bvec_0$. In order to avoid difficulties in defining
the term ``instability'' and to facilitate the calculations we assume $\Bvec_0$
to be constant in time. Consequently we are forced to assume
the existence of an additional e.m.f. which prevents
$\Bvec_0$ from decaying. Although appearing to be very artificial, we find
this measure to be legitimate as long as the results of the
stability analysis are applied to real physical situations obeying the
constraint that the background field $\Bvec_0$ is changing only slightly
during the considered period of time.

Linearization of \eqref{indeqdimless} about $\Bvec_0$ yields
\begin{equation}
\dot{\bvec} = \Delta\bvec - \curl (\,\curl\Bvec_0
\times\bvec\ + \curl\bvec\times\Bvec_0\,)\;,\;\;\dv \bvec = 0 
\label{indeqdimlesslin}
\end{equation}
describing the behavior of small perturbations $\bvec$ of the reference
state.

With respect to the magnetic energy balance Eq. \eqref{indeqdimlesslin} shows
a remarkable difference to Eq. \eqref{indeqdimless}. 
Along with the term $\curl\bvec\times\Bvec_0$ which is again
energy--conserving now as a second Hall--term $\curl\Bvec_0\times\bvec$ occurs
which may well deliver or consume energy (from/to $\bvec$ !) since in general the integral 
$\int_V (\curl\Bvec_0\times\bvec)\cdot \curl\bvec\, dV$ will not vanish.
This
reflects the fact that the linearized Hall--induction equation describes 
the behavior of only a part of the total magnetic field. Actually, perturbations
may grow
only on expense of the energy stored in the background field.
Considering \eqref{indeqdimlesslin}, we can determine a scale below
which the
ohmic dissipation dominates the Hall-drift. Estimating $|\curl\Bvec_0|$
and $|\curl\bvec\,|$ by $\bar B_0$ and $\bar b / l$, respectively,
we find the critical scale of $\bvec$ to be $l_{crit} \lesssim 1/\bar B_0$, which 
is de--normalized $l_{crit} \lesssim L/(\omega_{{\bar B}_0}\tau_e)$, identical with the 
expression derived in \cite{GR92} considering the Hall--cascade 
in analogy with the turbulent flow of an incompressible fluid.

Let us now specify the geometry of our model and 
the background field. We consider a slab which is infinitely extended
both into the $x$-- and $y$--directions but has a finite thickness $2L$ in
$z$--direction.
The background field is assumed to be
parallel to the surface of the slab pointing, say, in $x$--direction and to
depend on the $z$ coordinate only, i.e. $\Bvec_0 = f(z) \evec_x$. 
Guided by the conditions under which the above--mentioned
magnetic instability \cite{R69} may work, we conclude that 
$f(z)$ has to be at least quadratic thereby ensuring that the
first term in \eqref{indeqdimlesslin} is able to play the role of the
shear flow, the second the role of the $\omvec\times \jvec$--term.
Note, that by this choice $\curl\Bvec_0\times \Bvec_0$
represents a gradient. Thus the unperturbed evolution of the
background field is not at all affected by the Hall--drift; in the
absence of an e.m.f. it would decay purely ohmically!

Further on we decompose a perturbation $\bvec$ into a
poloidal and a toroidal component, $\bvec =  \bvec_p+\bvec_t$, which can
be  represented by scalar
functions $S$ and $T$, respectively, by virtue of the definitions
\begin{equation}
\hspace*{-7mm}\bvec_p = -\curl\,(\,\ezvec \times \nabla S)~~,\; 
\bvec_t = -\ezvec \times \nabla T\;, 
\hspace*{-1cm}\label{poltorST}
\end{equation}
ensuring $\dv\bvec=0$ for arbitrary $S,T$.

For the sake of simplicity we will confine ourselves to the
study of plane wave solutions with respect to the $x$-- and $y$--directions,
thus making the ansatz
\begin{equation}
\left\{\begin{aligned}
&S\\
&T
\end{aligned}\,\right\}(\xvec,\tau) = 
\left\{\begin{aligned}
&s\\
&t
\end{aligned}\,\right\} (z) \exp{(i\tilde{\kvec}\tilde{\xvec} + p\tau)}~~,
\label{transf}
\end{equation}
where $\tilde{\kvec} = (k_x,  k_y)$, $\tilde{\xvec} = (x,y)$
and $p$ is a complex time increment.
It guarantees as well the uniqueness of the poloidal--toroidal decomposition
since from $\undertilde{\lapl}(S,T\,)=0$ it follows $(S,T\,)=0$ with
$\undertilde{\lapl}$ being the 2--dimensional lateral Laplacian
(see \cite{BR88}).
With \eqref{transf} we obtain from \eqref{indeqdimlesslin} two coupled ordinary
differential equations
\begin{equation}
\hspace*{-1.1cm}\begin{alignedat}{4}
&pt -t'' &&+ \tilde{k}^2t &&= &&i k_x f (s'' -\tilde{k}^2s) - i k_x f''s\\ 
&ps -s'' &&+ \tilde{k}^2s - i k_y  f's &&=  - &&i k_x f t \;,
\end{alignedat}
\hspace*{-9mm}\label{poltoreq}
\end{equation}
where the dash denotes the derivative with respect to $z$. 
Together with appropriate boundary conditions Eqs. \eqref{poltoreq} define an
eigenvalue problem with respect to $p$.

We consider two types of boundary conditions (BC):
For the vacuum condition we assume $\curl \Bvec = \zervec$ outside the slab and
require continuity of all components of $\Bvec$ across the boundary.
For the perfect--conductor condition an electric field must be prevented from
penetrating into the region outside the slab, that is, the normal magnetic and
tangential electric field components must vanish 
at the boundary. In terms of the scalars $s$ and $t$ this means
$[s]=\left[s'\right]=t=0$
for the vacuum condition and
$s=t'=0$
for the perfect conductor condition where $[.]$ denotes the jump across a
boundary. For $t'=0$ to be valid the vanishing of $\Bvec_0$ at the
boundary is required.
Making
use of the vacuum solutions vanishing at infinity for either halfspace, 
$z\ge 1$ and $z\le -1$, respectively, the vacuum boundary condition for $s$ can be
expressed as $s' = \mp \tilde k s$ for $z=\pm 1$, with $\tilde k = |\tilde{\kvec}|$.

Obviously, three distinguishable combinations of the boundary conditions are
possible:
vacuum on either side (VV), perfect conductor on either side (PP), vacuum on
one
and perfect conductor on the other side of the slab (PV). The latter choice
comes closest to neutron star conditions if we think of the crust being
neighboured upon a
superconducting core on the one and a region with very low conductivity on the
other side. The VV boundary condition may in turn be appropriate for a galactic
disc. 
Since both the PV
and the VV BC were to be considered, we choose sufficiently curved background field
profiles, which obey them, i.e.
$f(z)=B_0(1+z)(1-z^2)$ and $f(z)=B_0(1-z^2)$ for BC=PV and BC=VV , respectively.

For certain ranges of the wave numbers ${k_x,k_y}$ and for $B_0 \gtrsim 3$ we
found eigenvalues $p$ with a positive real part, i.e. exponentially growing
perturbations.
The dependence of
the growth rate $\Re(p)$ on the wave numbers for $B_0=1000$ and BC$\,=\,$PV
is shown in Fig. \ref{fig1}.

Figure \ref{fig2} shows the dependence of growth rate and wave number $k_x$
of the fastest growing mode on $B_0$.

An interesting feature is, that the maximum growth rates occur
for all $B_0$ considered at $k_y=0$.
Of course this asymmetry is due to the choice of the
background field: once it was chosen parallel to the $y$--direction the maximum
growth rates would occur at
$k_x=0$. Another interesting result is the dependence of the growth rate on the
boundary conditions. BC$\,=\,$PV
yields the largest values, by a factor $1.6 \dots 3$ larger than for
BC$\,=\,$VV, while  
BC$\,=\,$PP results in very small growth rates. Note that the
most unstable eigenmodes are always non--oscillatory, though oscillating
unstable ones exist.

Evidently, the obtained growth rates are in agreement with
the constraint, formulated above: In comparison with the background field decay
the growth
of the most unstable perturbations is a fast process;
thus we may consider it as `episodically unstable'.

With respect to the asymptotic behavior $\sigma\!\rightarrow\!\infty$ for a fixed
(unnormalized!) background field 
one has to note
that the time increment $p$ is normalized on the ohmic decay rate ($\propto \sigma^{-1}$).
From Figure \ref{fig2} it can be inferred $\Re(p)\propto B_0^q\,,\;q < 1$ for $B_0 \ge 100$,
which means that in the limit of negligible
dissipation the growth rate in physical units tends to zero.    


Figure \ref{fig5} shows the eigensolutions $(s,t\,)(z)$ of the fastest growing
mode for three different values of $B_0$ and BC$\,=\,$PV.
One can observe that with increasing
$B_0$ the toroidal field
becomes more and more small--scaled and concentrated towards the vacuum
boundary. In contrast, the corresponding poloidal field remains large--scaled.

The magnetic field
structure of the fastest growing mode for $B_0=2000$ and BC$\,=\,$PV is shown
in Fig. \ref{fig6}.

Clearly, any assignment of the results gained by help of a very simplified
model to astrophysical objects has to be done with great care. Even when 
accepting the plane layer as a reasonable approximation of a neutron star's
crust
one has to concede that the very specific profiles of $\Bvec_0$ assumed above
may only exemplify the field structure in the crust. 

An acceptable approximation of the
radial
profile of a dipolar crustal field as given e.g. in \cite{PGZ00} will in
general have to allow for a
linear part and non--zero values at the boundaries. Moreover, the strong 
dependence of the conductive properties on the radial co--ordiante should anyway be
taken into account.


To get an impression of possible consequences for the evolution of neutron
stars
we now simply assume, that the real $\Bvec$--profile is sufficiently ``curved"
(i.e. its second derivative is big enough) and associate the parameter $B_0$
with a typical value of the field.

Assuming further electric conductivity and chemical composition to be constant,
$\sigma= 5\times10^{26}$s$^{-1}$ and the relative atomic weight $A/Z=25$, respectively,
we find the normalization field at a density $\rho = 10^{14}$g cm$^{-3}$
to be $7\times 10^{10}$G (see e.g. \cite{PGZ00}). That is, for typical (inner) crustal
magnetic 
fields ranging between $7\times 10^{12}$G and $1.4\times 10^{14}$G we find 
a $B_0$ between
$100$ and $2000$ and the e--folding time of the most rapidly growing unstable
mode to be $0.0035$ and $0.0003$ times the Ohmic decay time, respectively. 
Thus, an initial perturbation will 
quickly evolve to a level at which the linear analysis is no longer feasible,
that is, at which it starts to drain a remarkable amount of energy out of the
background field.
 
We want to emphasize again that a sufficient curvature of the background field
profile
is a necessary condition for the occurrence of an unstable behavior. Therefore 
neither the derivation of the well--known helicoidal waves (whistlers) nor its
modification presented in \cite{VCO99} could reveal it because a homogeneous
background field is used.
 
With even more care we may speculate about possible observational consequences.
The instability discussed here may perhaps act effectively in the
crust of not too young (age $t \gtrapprox 10^5$ yrs) neutron stars. 
For those stars, the small--scale field modes initially generated or existing
in the crust have already been decayed and the
magnetic field is concentrated almost completely in the large--scale, say,
dipolar mode.
Simultaneously, in the process of cooling the coefficient $m_e^* c/ e\tau_e$
becomes smaller and smaller until the nonlinear Hall term in
\eqref{indeqdimless}
dominates the linear ohmic term. From that moment the Hall--instability may
rise small--scale modes down to scale lengths $\gtrsim l_{crit}$ on expense of
the dipolar mode. This would lead to a change of the
spin--down behaviour of isolated neutron stars. Deviations from the
``standard'' rotational evolution will occur
when the dipolar field decreases rapidly due to the instability. This
may lead to the observation of braking indices $n > 3$ \cite{JG99} during the action
of that instability. Another possible
observational consequence is due to enhanced Joule
heating,
which will keep the neutron star warmer than standard cooling
calculations predict after an age critical for the onset of the
Hall--instability ($\gtrapprox 10^5$ yrs). Third, the strong small-scale field
components cause strong small--scale Lorentz forces which may be able to crack
the crust. This could be observable in glitches or, depending on the available
energy even in Gamma-- and X--ray bursts \cite{TD95}.

\vspace{.1cm}
The authors are grateful to K.--H. R\"adler, D. Konenkov and K. Baumg\"artel (AIP),
K. G\"artner (Weierstrass Inst. Berlin), A. Reisenegger (PUC Chile) and
D. Yakovlev (Joffe Inst. St. Petersburg) for helpful discussions.

\pagebreak
\begin{figure}
\caption{Growth rate as a function of $k_x$ and $k_y$ for
$B_0=1000$ and BC$\,=\,$PV. Negative values were set to zero.}
\label{fig1}
\end{figure}

\begin{figure}[t]

\caption{Growth rate and wave number $k_x^\tmax$ of the fastest growing mode as
functions of $B_0$.
Solid and dashed lines correspond to BC$\,=\,$PV and BC$\,=\,$VV, while thick
and thin lines correspond to growth rates and $k_x$, respectively.}
\label{fig2}
\end{figure}

\begin{figure}[h]

\caption{Moduli of $(s,t\,)(z)$ of the fastest growing mode; BC$\,=\,$PV.
Solid, dash--dotted and dashed lines refer to $B_0=2000$, $B_0=100$ and
$B_0=10$, respectively.}
\label{fig5}
\end{figure}

\begin{figure}[t]
\begin{center}
\end{center}

\caption{Field structure of the fastest growing mode for $B_0=2000$ and
BC$\,=\,$PV.
Arrows: $b_{x,z}$, grey shading: value of $b_y$, dark --- into, light --- out
of the plane.}
\label{fig6}
\end{figure}

\begin{thebibliography}{}

\vspace{-5mm}
\bibitem{YS91}
Yakovlev, D. and Shalybkov D., Astrophys. Space Sc. {\bf 176}, 191 (1991)

\bibitem{HUY90SU97}
Haensel, P., Urpin, V., and Yakovlev, D., Astron. Astrophys. {\bf 229}, 133
(1990); Shalybkov, D. and Urpin, V., Astron. Astrophys. {\bf 321}, 685 (1997)

\bibitem{GR92}
Goldreich, P. and Reisenegger, A., Astrophys. J. {\bf 395}, 250 (1992)

\bibitem{M94}
Muslimov, A.G., Mon. Not. R. Astron. Soc. {\bf 267}, 523 (1994)

\bibitem{NK94US99}
Naito, T. and Kojima, Y., Mon. Not. R. Astron. Soc. {\bf 266}, 597 (1994);
Urpin, V. and Shalybkov D., Mon. Not. R. Astron. Soc. {\bf 304}, 451 (1999)

\bibitem{SU95}
Shalybkov, D. and Urpin, V., Astron. Astrophys. {\bf 273}, 643 (1995)

\bibitem{US95}
Urpin, V. and Shalybkov D., Mon. Not. R. Astron. Soc. {\bf 294}, 117 (1995)

\bibitem{VCO99}
Vainshtein S., Chitre S., and Olinto A., Phys. Rev. E {\bf61}, 4422 (2000)

\bibitem{WG96}
Wiebicke,  H.--J. and Geppert, U., Astron. Astrophys. {\bf 309}, 203 (1996)

\bibitem{R69}
R\"adler, K.--H., Monatsber. Dt. Akad. Wiss. {\bf 11}, 194 (1969)

\bibitem{M78}
Moffat, H. K., {\it Magnetic field generation in electrically conducting
fluids.} Cambridge University Press 1978

\bibitem{BR88}
Br\"auer, H.--J. and R\"adler, K.--H., Astron. Nachr. {\bf 309}, 1 (1988)

\bibitem{PGZ00}
Page D., Geppert U., and Zannias T., Astron. Astrophys. {\bf 360}, 1052 (2000)

\bibitem{JG99}
Johnston, S. and Galloway, D., Mon. Not. R. Astron. Soc. {\bf 306}, L50 (1999)
 
\bibitem{TD95}
Thompson C. and Duncan R., Astrophys. J. {\bf 473}, 322 (1995)

\end{thebibliography}
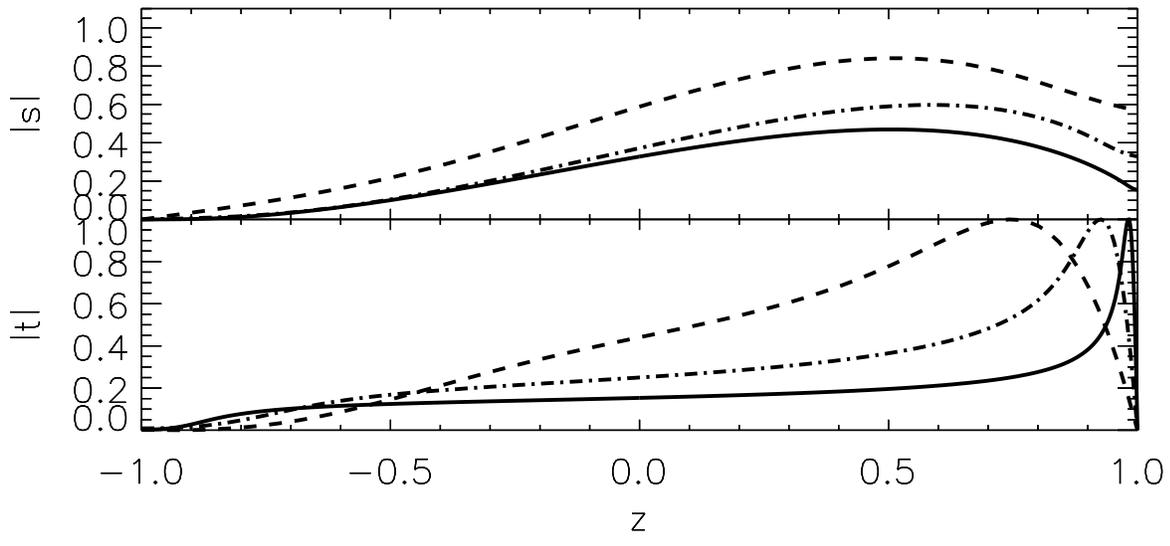
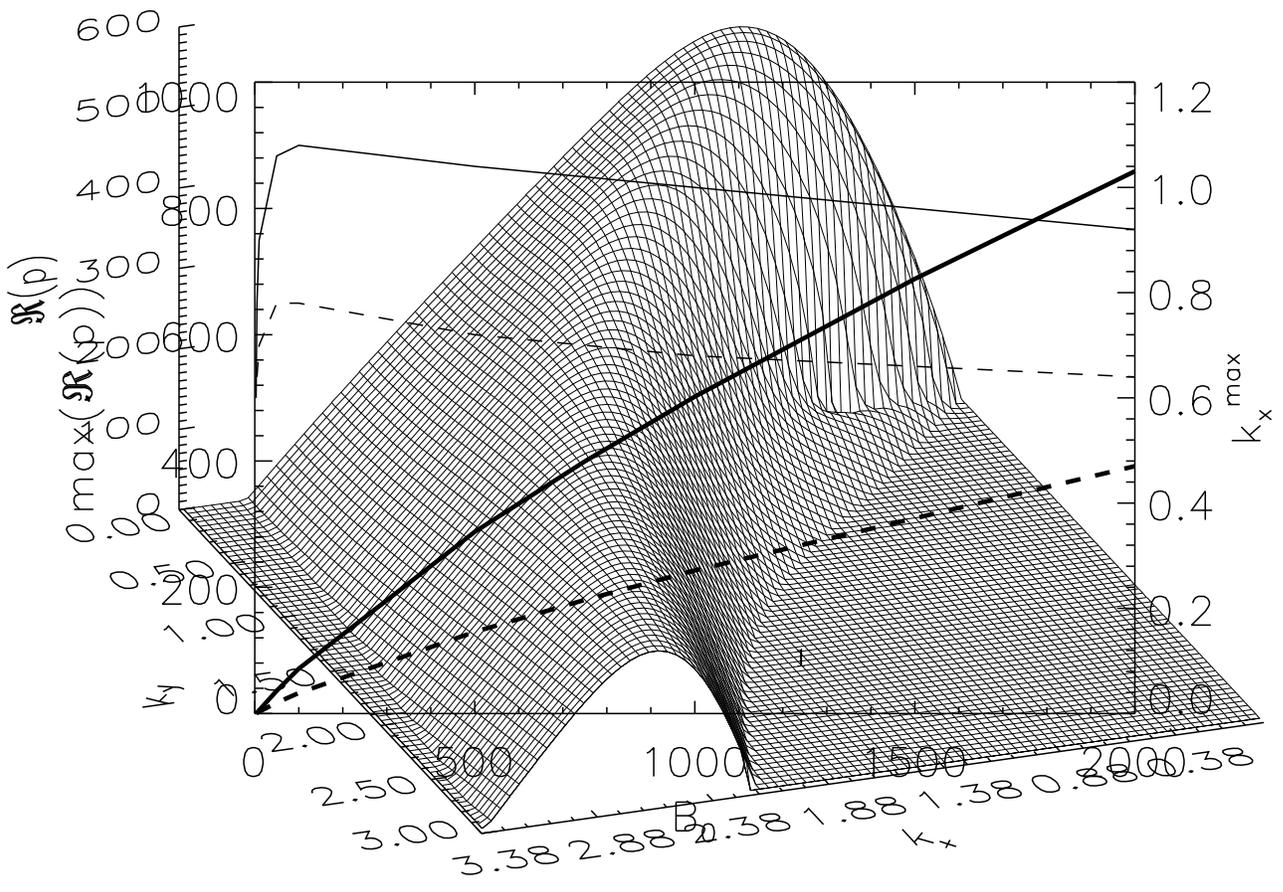
\end{document}